\documentclass[letter]{aa}  
\usepackage{graphicx}
\usepackage{CJKutf8}
\usepackage[varg]{txfonts}
\usepackage{amsmath}	
\usepackage{mathrsfs}
\usepackage{bm}
\usepackage{color}
\usepackage{ulem}
\usepackage{xspace}
\usepackage{makecell}
\usepackage{threeparttable}
\usepackage{hyperref}
\hypersetup{colorlinks=true, linkcolor=blue, citecolor=blue, urlcolor=blue}
\usepackage{orcidlink}
\usepackage{txfonts}
\usepackage[version=4]{mhchem}

\definecolor{mygray}{gray}{0.6}
\definecolor{orange}{rgb}{1.0, 0.4, 0.0}
\definecolor{myblue}{rgb}{0.1, 0.5, 0.7}
\definecolor{mygreen}{rgb}{0.2, 0.6, 0.4}
\newcommand{\ccc}[1]{\textcolor{orange}{[\textit{CWO: \small #1}]}}

\newcommand{\corem}[1]{\textcolor{mygray}{\sout{#1}}}
\newcommand{\cim}[1]{\textcolor{orange}{[\textit{$\star$chris$\star$}:\textbf{\small #1}}]}

\newcommand{\md}{\textbf}
\newcommand{\mdd}{\textbf}

\definecolor{mygray}{gray}{0.6}
\newcommand{\ywc}[1]{[\textcolor[RGB]{128,0,128}{\small YW: \textit{#1}}]} 
\newcommand{\ywrem}[1]{\textcolor{mygray}{\sout{#1}}} 

\newcommand{\app}[1]{App.~\ref{sec:#1}}

\newcommand{\se}[1]{Sect.~\ref{sec:#1}}

\newcommand{\fg}[1]{Fig.~\ref{fig:#1}}
\newcommand{\fgs}[2]{Figs.~\ref{fig:#1} and \ref{fig:#2}}
\newcommand{\tb}[1]{Table~\ref{tab:#1}}

\newcommand{\eq}[1]{equation~(\ref{eq:#1})}

\newcommand{\eqs}[2]{equations~(\ref{eq:#1}) and (\ref{eq:#2})}

\newcommand\gas{\mathrm{g}}
\newcommand\dust{\mathrm{d}}
\newcommand\parti{\mathrm{p}}
\newcommand\diff{\mathrm{d}}

\iftrue
\renewcommand{\ywc}[1]{}
\renewcommand{\ywrem}[1]{\xspace}

\renewcommand{\cim}[1]{}

\renewcommand{\corem}[1]{\xspace}
\renewcommand{\ccc}[1]{\xspace}

\renewcommand{\md}{}
\renewcommand{\mdd}{}
\fi

\begin{document} 


   \title{Dust and Water in V883 Ori: Relics of a Retreating Snowline} 

   \author{
            Y.~Wang (\begin{CJK*}{UTF8}{gbsn}王雨\end{CJK*})
          \inst{1} \orcidlink{0009-0004-2217-4439}
          \and
          C.W.~Ormel\inst{1} \orcidlink{0000-0003-4672-8411}
          \and
          H.-C. Jiang (\begin{CJK*}{UTF8}{gbsn}蒋昊昌\end{CJK*})\inst{2} \orcidlink{0000-0003-2948-5614}
          \and
          S.~Krijt\inst{3} \orcidlink{0000-0002-3291-6887}
          \and
          A.~Houge\inst{4} \orcidlink{0000-0001-8790-9011}
          \and
          E.~Macías\inst{5} \orcidlink{0000-0003-1283-6262}
          }

        \institute{Department of Astronomy, Tsinghua University, 30 Shuangqing Rd, Haidian DS, 100084 Beijing, China\\
	\email{wang-y21@mails.tsinghua.edu.cn}
        \and
            Max Planck Institute for Astronomy, Königstuhl 17, 69117 Heidelberg, Germany
        \and
            Department of Physics and Astronomy, University of Exeter, Exeter, EX4 4QL, UK
         \and
            Center for Star and Planet Formation, GLOBE Institute, University of Copenhagen, Øster Voldgade 5-7, DK-1350 Copenhagen, Denmark
        \and 
            European Southern Observatory, Karl-Schwarzschild-Str 2, 85748 Garching, Germany}

   \date{Received September 15, 1996; accepted March 16, 1997}

 
\abstract{
    V883 Ori is an FU-Orionis-type outburst system characterized by a shoulder at 50-70 au in its ALMA band 6 and 7 intensity profiles. Previously, this feature was attributed to dust pile-up from pebble disintegration at the water snowline. However, recent multi-wavelength observations show continuity in the spectral index across the expected snowline region, disfavoring abrupt changes in grain properties. Moreover, extended water emission is detected beyond 80 au, pointing to a snowline further out. This Letter aims to explain both features with a model in which the snowline is receding.
    We construct a 2D disk model that solves the cooling and subsequent vapor recondensation during the post-outburst dimming phase.
    Our results show that both the intensity shoulder and the extended water emission are natural relics of a retreating snowline: the shoulder arises from excess surface density generated by vapor recondensation at the moving condensation front, while the outer water vapor reservoir persists due to the long recondensation timescales of $10^{2}-10^{3}$\ yr at the disk atmosphere.
    As V883 Ori continues to fade, we predict that the intensity shoulder will migrate inward by an observationally significant amount of 10 au over about \md{25} years.}

   \keywords{Protoplanetary disks --
                Planets and satellites: formation --
                Stars: protostars
               }

   \maketitle
%
\section{Introduction}
In protoplanetary disks, snowlines mark the locations where volatile ice sublimates. This process dictates volatile distributions and alters grain properties in the disk (e.g., \citealt{OebergEtal2023}).
In this context, outburst systems are ideal laboratories for studying the effect of snowlines. Due to an abrupt increase in accretion, these young stellar objects brighten by many orders of magnitude, heating up the disk for tens to hundreds of years and enabling large-scale sublimation of ices (e.g., \citealt{FischerEtal2023}). In particular, the snowline is pushed to $\sim$10--100 au, bringing it within reach of facilities such as the Atacama Large Millimeter/submillimeter Array (ALMA).

V883 Ori is one of the best-studied FU Orionis outburst systems. \citet{CiezaEtal2016} identified a ``dark annulus'' in ALMA band 6 continuum,  \md{suggesting} that its water snowline lies at $\approx$40 au. This shoulder-like feature in the intensity profile was initially attributed to the accumulation of small grains inside the snowline, produced by the \md{disintegration} of dust grains after ice sublimation \citep{SaitoSirono2011, AumatellWurm2011}.
However, multi-wavelength analysis has revealed \md{continuity in the} spectral index across the intensity shoulder, implying no abrupt change of the grain physical properties by ice sublimation \citep{HougeKrijt2023,HougeEtal2024}. 
The exact snowline location is further complicated by the detection of H$_{2}^{18}$O \citep{TobinEtal2023} and indirect tracers like methanol or HCO$^{+}$ \citep{van'tHoffEtal2018, LeemkerEtal2021}, which suggest a snowline location ${\gtrsim}80\,\mathrm{au}$.

In this Letter, we demonstrate that these observations can be consistently explained by a model in which the snowline retreats \md{along} with the decline of bolometric luminosity. At the \md{inward-}moving snowline front, the intensity shoulder \md{arises from} the \md{increase in} surface density of solids due to ice recondensation, while in the outer disk the sluggish recondensation of vapor preserves a relic of a previously hot state.




\section{Model}
\label{sec:model}
We set up the disk at the outburst active phase, when the bolometric luminosity reaches its peak value (see \se{dim_curve}), assuming the dust-vapor mixture is in hydrostatic balance and thermal equilibrium. The main focus of this work is to model the subsequent dimming phase, when $L_{\mathrm{bol}}$ declines and vapor recondenses onto grain surfaces. A key assumption is that dimming proceeds rapidly (${\sim}200~\mathrm{yr}$; see \se{dim_curve}) so that dynamic processes (gas advection, pebble drift, and diffusion) can be neglected.

\subsection{Disk structure}
\label{sec:disk_model}
The gas surface density profile is taken as a power law
\begin{equation}
\label{eq:sigma_gas}
    \Sigma_{\gas} (r) = \Sigma_{c} \left( \frac{r}{r_{c}}\right)^{-\gamma}.
\end{equation}
Following the fit of dust thermal emission by \citet{CiezaEtal2018}, \md{we take $r_{c} = 31\,\mathrm{au}$ and $\Sigma_{c}=160~\mathrm{g~cm^{-2}}$, while $\gamma$ is a treated as a parameter given its poor constraint \citep{van'tHoffEtal2018}}. The surface density of pebbles is taken as:
\begin{equation}
\label{eq:sigma_peb}
    \Sigma_{\parti}= Z_{\mathrm{peb}} \Sigma_{\gas}(r_{c}) \left( \frac{r}{r_{c}}\right)^{-\gamma_{\mathrm{peb}}},
\end{equation}
where $Z_\mathrm{peb}$ is the metallicity at the reference position. 
Beyond the snowline, pebbles are \md{assumed} 50\% \md{refractory} and 50\% water ice by mass, \mdd{with the material densities of constituent species following DSHARP \citep{BirnstielEtal2018}}.

Following \citet{WangEtal2025} we construct a 2D model in the disk's $R{-}z$ plane, where a standard viscous disk with $\alpha=10^{-3}$ \citep{ShakuraSunyaev1973} is assumed. For solids, we adopt an MRN size distribution \citep{MathisEtal1977}, which is appropriate when dust growth is limited by fragmentation \citep{BirnstielEtal2011}. The size distribution is characterized by an upper bound $s_{\max}$, corresponding to a midplane Stokes number $\mathrm{St}_{0}$, a lower bound of $s_{\min}=0.1\,\mu\mathrm{m}$, and is assumed unchanged by the outburst \citep{HougeEtal2024}.
Pebbles of different sizes are vertically distributed according to their scale height $H_{\mathrm{p},i}(H_\mathrm{g}, \mathrm{St}_{i,0},\alpha)$, where $H_{\gas}$ is the gas pressure scale height and $\mathrm{St}_{i,0}$ the Stokes number of pebble $i$ at the midplane (see \app{dust_bins}).

The vapor condensation rate ($\mathcal{R}_{\mathrm{cond}}$) is determined by its saturation condition and the total surface area of particles (e.g., \citealt{RosJohansen2013, SchoonenbergOrmel2017}). Specifically, the vapor density at any point in the disk reduces at a rate,
\begin{equation}
\label{eq:limiter}
    -\frac{\diff \rho_\mathrm{vap}}{\diff t} 
    = \mathcal{R}_{\rm cond} 
    = \langle \pi s^{2}_{\parti} n_\parti \rangle v_{\mathrm{th,vap}} \rho_{\mathrm{vap}} \left(1 - \frac{P_{\mathrm{eq}}}{P_{\mathrm{vap}}}\right).
\end{equation}
where $\langle \pi s^{2}_{\parti} n_\parti \rangle$ is the surface area of pebbles per unit volume (\app{dust_bins}, \eq{surface_area}).

\subsection{Luminosity during outburst dimming phase}
\label{sec:dim_curve}
\begin{figure}
    \centering
    \includegraphics[width=\columnwidth]{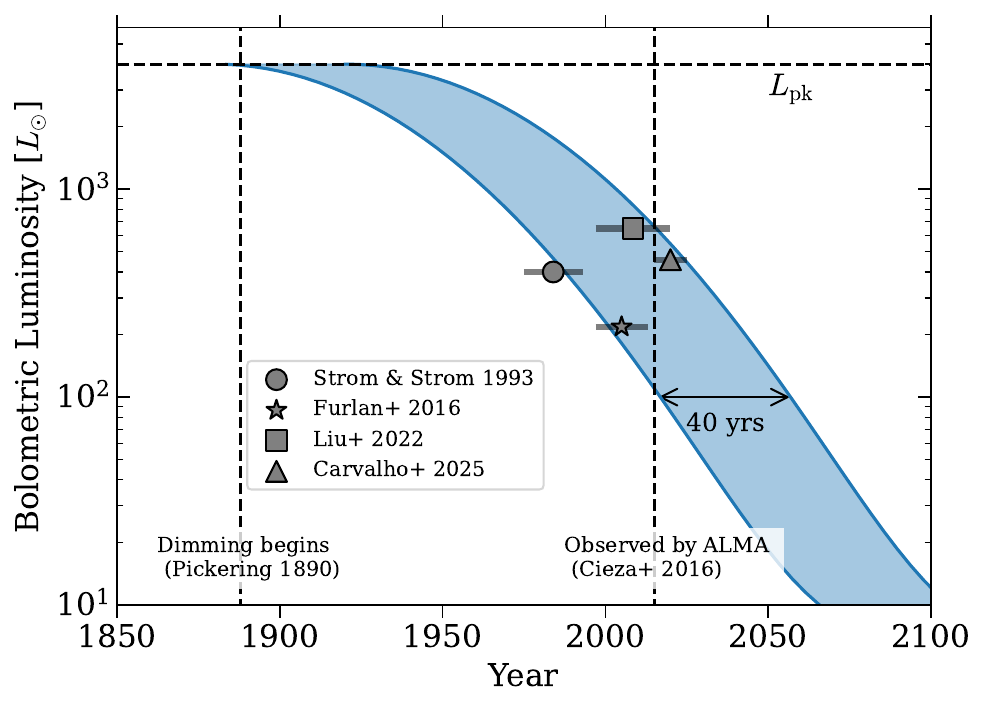}
    \caption{Evolution of the bolometric luminosity of V883 Ori during the outburst dimming phase. The symbols denote measurements from different studies, \md{with horizontal bars showing the time range of adopted data}. The blue band represents the adopted luminosity curve, which carries an uncertainty of ${\sim}40~\mathrm{yrs}$ due to the uncertainties in estimating $L_{\mathrm{bol}}$.}
    \label{fig:dimming_curve}
\end{figure}

We model the dimming of V883 Ori following the published luminosities, accounting for observational uncertainties arising from dust extinction and instrumental saturation (see \app{L_bol_uncertain}). Specifically, the time-dependent bolometric luminosity reads,
\begin{equation}
    \label{eq:Lfunc}
    L_{\mathrm{bol}} (t) = L_{\star} + (L_{\mathrm{pk}} - L_{\star}) \exp\left[-0.5\left(\frac{t-t_{\mathrm{beg}}}{50~\mathrm{yr}}\right)^{2} \right],
\end{equation}
where $L_{\star}$ is the stellar luminosity ($6\,L_{\odot}$; \citealt{CiezaEtal2016}). We adopt a peak luminosity of $L_{\mathrm{pk}}=4\,000\,L_{\odot}$ (comparable to that used in \citealt{LeemkerEtal2021}), which pushes the water snowline to 80--100~au. The parameter $t_{\mathrm{beg}}$ indicates the year when the outburst begins to fade.  As shown in \fg{dimming_curve}, choosing $t_{\mathrm{beg}}$=\md{1880} A.D. makes the dimming curve consistent with the lower limit of the estimated $L_{\mathrm{bol}}$ of V883 Ori (400 $L_{\odot}$, \citealt{StromStrom1993}; and 218 $L_{\odot}$, \citealt{FurlanEtal2016}). By shifting $t_{\mathrm{beg}}$ forward by $40\,\mathrm{yr}$ (A.D.~1920), the curve reaches the upper limit (647 $L_{\odot}$) given by \citet{LiuEtal2022}. Recently, \citet{CarvalhoEtal2025} fit the spectrum at 0.4-4.2 $\mu \mathrm{m}$ and obtained an accretion luminosity of 458 $L_{\odot}$, lying well within the assumed range. Through \eq{Lfunc}, the model output is only a function of $t'=t-t_\mathrm{beg}$, the time since $L_{\mathrm{bol}}$ starts to decline. 

\subsection{Temperature model}
\label{sec:T_model}
We adopt a two-stream radiation transfer (2sRT) method to calculate the temperature (e.g., \citealt{Hubeny1990}).
The heating sources at each grid cell are due to irradiation ($q_{\mathrm{irr}}$), viscous dissipation ($q_\mathrm{vis}$) and latent heat exchange ($q_{\mathrm{latent}}$).  
We compute $q_{\mathrm{irr}}$ from the bolometric luminosity (\eq{Lfunc}), while the viscous heating rate follows from the $\alpha$-disk viscosity \md{($q_{\mathrm{vis}}=\frac{9}{4} \rho \alpha c_{\mathrm{s}}^{2} \Omega$)} and the latent heating rate is proportional to the \md{condensation} rate (\eq{limiter}, see \citealt{WangEtal2025}).
In the outburst active phase, the disk adjusts to a temperature and density structure determined by the bolometric luminosity $L_\mathrm{pk}$, assuming hydrostatic balance.
In the dimming phase, following \citet{FlockEtal2017}, we let the disk adapt to a new equilibrium temperature obtained by the 2sRT ($T_{\mathrm{eq}}$) over a thermal relexation timescale, $\mathrm{d} T /\mathrm{d}t = (T_{\mathrm{eq}} - T) /t_{\mathrm{relax}}$ (see \app{t_relax} for the calculation of $t_{\mathrm{relax}}$). The Rosseland mean opacity per unit gas mass $\kappa_{\mathrm{R}}$ -- a key quantity that determines $t_{\mathrm{relax}}$ -- is a free model parameter. In our simulation (see \se{results}), due to the low metallicity and disk viscosity, the disk temperature is \md{determined} by irradiation, with $q_\mathrm{vis}$ and $q_{\mathrm{latent}}$ contributing little.

\subsection{Fitting ALMA observations}
\label{sec:fit_alma}
The intensity shoulder in V883 Ori is identified in both ALMA band 6 ($0.038''$, project code: 2015.1.00350.S, PI: Lucas Cieza) and band 7 ($0.028''$, project code: 2016.1.00728.S, PI: Lucas Cieza) high-resolution continuum images.
\md{We converted the continuum images into radial profiles by azimuthally averaging the data within concentric annuli of radial width 1/5 of the respective beam size. To fit the continuum profiles,} we generate synthetic intensity profiles in both bands by conducting 2sRT with \md{profiles for the} dust temperature and density taken from the simulation results. The 2sRT is performed in plane-parallel manner along the disk's vertical direction, with the optical depth corrected for disk inclination $i=38.3^{\circ}$ \citep{CiezaEtal2016}. Pebble opacities for different sizes and ice fractions are computed with \texttt{optool} \citep{DominikEtal2021}, adopting the DSHARP optical constants \citep{BirnstielEtal2018}. \md{After being convolved with the beam size, the synthetic intensity profiles are compared with the continuum data at 45–85 au, where the intensity shoulder lies}.


\citet{TobinEtal2023} measured the column density of water isotope H$_{2}^{18}$O in band 5 ($0.126''$). Given the limited data sensitivity, we fit only the mean column density between 80–120 au in our simulations to the observed value, assuming an isotopic ratio $^{18}$O/$^{16}$O = 1/560 \citep{WilsonRood1994}. In reality, the observed water vapor abundance may be influenced by processes beyond simple recondensation (\se{water_emission}). Therefore, we introduce an additional free parameter $f_{\mathrm{cond}}$ such that, $- \dust \rho_{\mathrm{vap}} / \dust t = f_{\mathrm{cond}} \mathcal{R}_{\mathrm{cond}}$.

To incorporate observational constraints from both continuum and water emission, we apply a \md{log-normal} likelihood for each dataset, \md{and} weight each likelihood with \md{the number of beams covered by the area of interest} (see \eqs{likelihood_i}{likelihood}).

%

\section{Results}
\label{sec:results}

\renewcommand{\arraystretch}{1.2}
\begin{table}[]
    \centering
    \caption{Overview of MCMC parameters and posterior values \md{([16,50,84]-th percentiles)} for both the retreating snowline and static snowline model.}
    \begin{tabular}{c|l|c|c}
        \hline
         & Prior & \multicolumn{2}{c}{\makecell{Posterior \\ retreating / static}} \\
        \hline
        $ \log Z_{\mathrm{peb}}$ & [-3.5, -2] & $-2.89^{+0.05}_{-0.05}$ & $-2.84^{+0.007}_{-0.007}$  \\
        $\gamma$ & [-2.0, -1.0] & $-1.31^{+0.22}_{-0.34}$  & $-1.99^{+0.005}_{-0.002}$ \\
        $\gamma_{\mathrm{peb}}$ & [-1.5, -0.5] & $-0.97^{+0.10}_{-0.10}$ & $-0.50^{+0.004}_{-0.008}$ \\
        $\log$ St$_{0}$ & [-3, -1] & $-1.81^{+0.13}_{-0.12}$ & $-2.71^{+0.016}_{-0.010}$ \\
        $t^{\prime}$ (yr) & [60, 170] & $120.8^{+9.3}_{-9.0}$ & -  \\
        $\log f_{\mathrm{cond}}$ & [-0.5, 1.5] &  $0.53^{+0.09}_{-0.08}$ & -\\
        $\log \kappa_{\mathrm{R}}$ & [0, 1.3] & $0.67^{+0.17}_{-0.15}$ & $0.006^{+0.001}_{-0.000}$ \\
        \hline
        $L_{\mathrm{bol}} (L_{\odot})$ & [0, 978] & - & $489.9^{+6.4}_{-6.0}$ \\
        \hline 
    \end{tabular}
    \tablefoot{$Z_{\mathrm{peb}}$: pebble metallicity (Eq.\ref{eq:sigma_peb}); $\gamma$, $\gamma_{\mathrm{peb}}$: gas and pebble surface density index (Eq.\ref{eq:sigma_gas},\ref{eq:sigma_peb}); St$_{0}$: Stokes number of maximum grain size (\se {disk_model}, \ref{sec:dust_bins}); $t^{\prime}$: time since the outburst dimming (Eq.\ref{eq:Lfunc}); 
    $f_{\mathrm{cond}}$: condensation rate correction factor (\se{fit_alma}); $\kappa_{\mathrm{R}}$: Rosseland-mean opacity (\se{T_model}, \ref{sec:t_relax}); $L_{\mathrm{bol}}$: bolometric luminosity, for static model (\se{static_model}).}
    \label{tab:mcmc_paras}
\end{table}
We conduct Markov chain Monte Carlo (MCMC) analysis with \textit{emcee} \citep{Foreman-MackeyEtal2013} to obtain the posterior distribution (see \se{MCMC_result}). The \md{parameter's} posterior values are given in \tb{mcmc_paras}. 

\subsection{Emergence of intensity shoulder}
\label{sec:intensity_shoulder}
The best-fit results are shown in \fg{obv_compare}, where the synthetic intensity profiles agree well with the dual-band data, especially capturing the intensity shoulder feature at 50-70 au (gray-shaded region). To understand this, we plot the time evolution of the corresponding surface density profiles in \fg{sigma_time}. As the disk cools during the outburst dimming phase, vapor gradually recondenses on pebbles, creating a bump in surface density at the condensation front, which appears in continuum as the intensity shoulder. Matching the observed position of the shoulder requires the snowline to retreat to $\approx$60 au at the observational epoch ($t'=121\,\mathrm{yr}$). This places a stringent constraint on the thermal relaxation time, or IR-opacity $\kappa_{\mathrm{R}}$, of the disk. If the disk had cooled more rapidly (lower $\kappa_R$), the snowline would already have retreated inside 50 au, shifting the shoulder much closer to the star, within the expected observational time.

Simultaneously fitting the dual-band continuum yields $\mathrm{St_{0}} \approx 0.01$, corresponding to a maximum grain size of ${\sim}$centimeter. The need of having cm-sized particles in the disk follows from the low spectral indices at 50-70 au ($\alpha=2.5-3.0$), a result consistently found across ALMA band 3-7 in V883 Ori by \citet{HougeEtal2024}.


\begin{figure}
    \centering
    \includegraphics[width=\columnwidth]{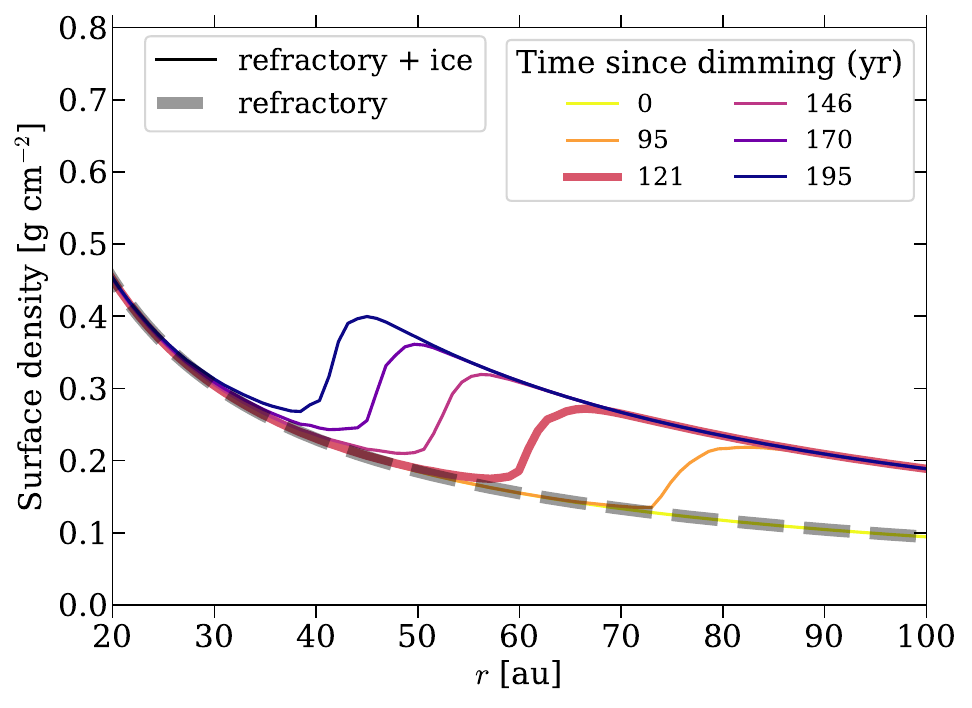}
    \caption{Time evolution of the surface density profiles of the best-fit model. Silicate surface density remains constant with time. The thick line denotes the best-fit $t^{\prime}=121$ yr.}
    \label{fig:sigma_time}
\end{figure}

\begin{figure}
    \centering
    \includegraphics[width=\columnwidth]{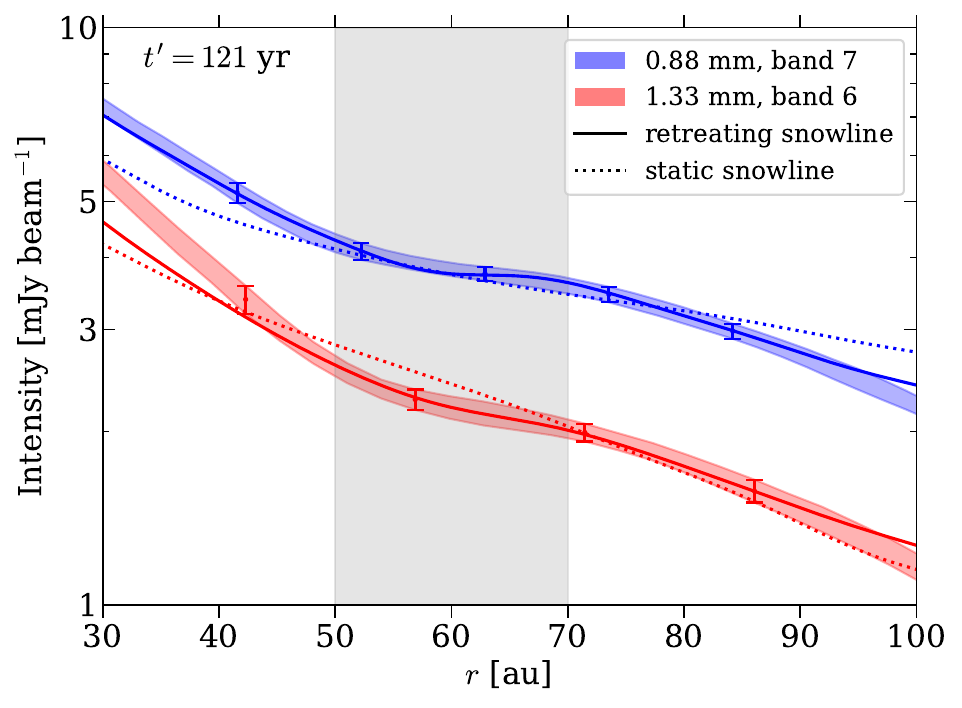}
    \caption{Comparison of the synthetic intensity profiles with ALMA continuum. Crosses show ALMA data adopted from \citet{HougeEtal2024}, spaced according to the beam sizes; shaded regions indicate the rms error. Solid lines represent the best-fit retreating snowline model, while dotted lines correspond to the best-fit static model. Enhanced emission towards the inner regions arises likely from intense viscous heating \citep{AlarconEtal2024}, which is not accounted for in the model.}
    \label{fig:obv_compare}
\end{figure}

\begin{figure}
    \centering
    \includegraphics[width=\columnwidth]{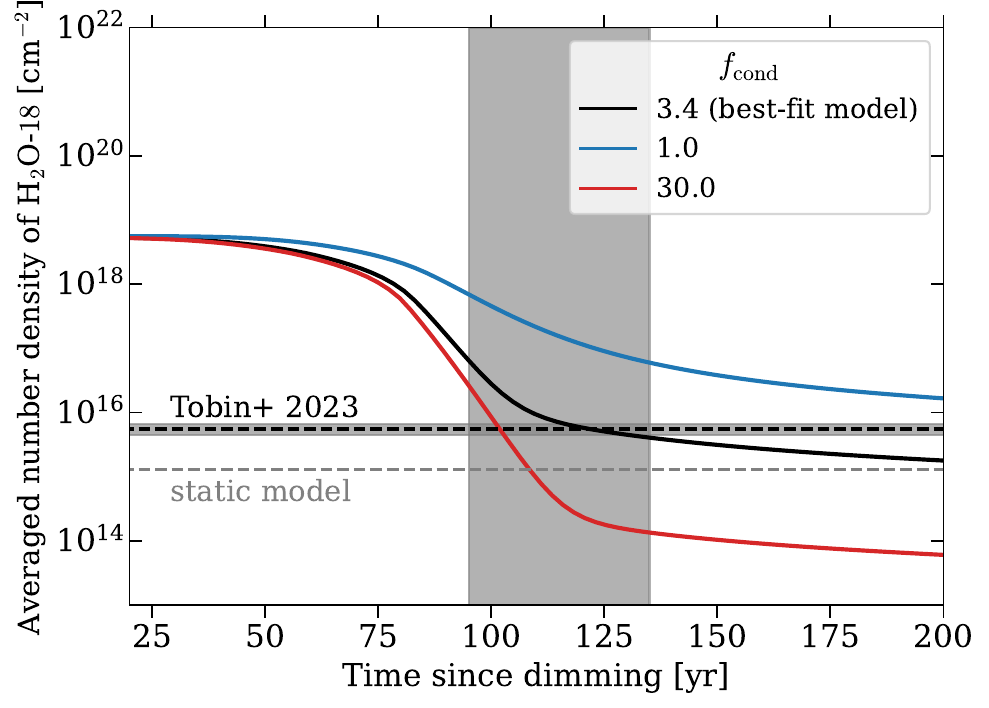}
    \caption{Average column density of H$_{2}^{18}$O at 80-120 au under different condensation rate ($f_{\mathrm{cond}}$). The vertical grey band indicates the time range when observing the intensity shoulder. The dashed \md{black} line denotes the average column density derived from ALMA band 5 data by \citet{TobinEtal2023} with $1\sigma$ uncertainty shaded in grey. \md{Dotted grey line denotes the best-fit value in static snowline model.}}
    \label{fig:water_sigma}
\end{figure}

\subsection{Extended water emission}
\label{sec:water_emission}


The mean column density of H$_{2}^{18}$O at 80-120 au, as estimated from our models assuming optically thin emission, is presented in \fg{water_sigma}.
In all simulated cases, the water abundance drops sharply within the first 110 years, and transitions to a slowly-evolving stage afterwards. This is due to different condensation timescales in the disk's midplane and the upper layers (see \app{2D_structure}). After the snowline retreats, vapor in the midplane recondenses rapidly. Conversely, in the disk's upper layer, where pebble densities are much lower, the recondensation time can reach ${\sim}10^2{-}10^3\,\mathrm{yr}$ \citep{VisserEtal2015,RabEtal2017}.
Therefore, due to the lag caused by the finite recondensation time, the snowline inferred from molecular emission would consistently appear further away than its actual position during the outburst dimming phase.
Our modeling shows that an enhanced condensation rate ($f_\mathrm{cond}\approx4$) is required to match the inferred water vapor abundance, which may indicate a shallower size distribution and/or porous grains \citep{Birnstiel2024}, both of which increase the total dust surface area at the disk atmosphere.
In addition, the destruction of water by photodissociation, heterogeneous nucleation and different outburst duration may also influence the observed gas-phase water abundance \citep{FischerEtal2023,RosJohansen2024,Romero-MirzaEtal2024i}.

\subsection{The need for a retreating snowline model}
\label{sec:static_model}
A retreating snowline naturally explains the distant intensity shoulder and the extended water emission, relics of the earlier outburst.
In contrast, a static snowline model fails to explain the observations (\fg{obv_compare} and \fg{corner_static}).
To illustrate this, we performed another MCMC simulation, where the disk structure is set by a constant $L_{\mathrm{bol}}$ (treated as a free parameter) rather than evolving with the luminosity curve. In other words, the temperature and vapor equilibrium instantly adapt to the stellar luminosity. 
\md{Because of the instantaneous adjustment, the static model fails to maintain a sufficient amount of vapor in the outer disk: the best-fit water number density is $\approx$$7 \sigma$ away from the observed value (\fg{water_sigma}). Furthermore, in the static model, the snowline is located at $\approx$30 au, failing to reproduce the shoulder feature at 50-70 au in the observed intensity profile.}
 
\section{Conclusions and discussions}
\label{sec:conclusion}
We investigated V883 Ori's continuum and water line emission by a retreating snowline model.
The key findings are:
   \begin{enumerate}
       \item The dimming stage of a stellar outburst results in a receding water snowline. Vapor condensation at the snowline front creates a bump in surface density, marking the transition from icy to ice-free pebbles.
       \item Vapor in the disk upper layer takes ${\sim}10^{2}{-}10^3\,\mathrm{yr}$ to recondense, leading to substantial amounts of water vapor in the outer disk, even after the snowline has retreated inwards. The observed H$_2^{18}$O outside the current snowline region (50-70 au) is a leftover of a much hotter past phase.
       \item The retreating snowline model simultaneously reproduces the intensity shoulder feature in ALMA band 6/7 continuum and the extended water emission. In contrast, a static model fails to maintain the snowline at 50-70 au under the present-day luminosity.
\end{enumerate}
In our modeling, both the disk cooling time ($t_{\mathrm{relax}}$), which controls the snowline retreat speed, and the mm opacity, which sets the continuum emission, depends on dust opacities. Thus, the thermal history of outburst systems constrains dust properties. For example, the retreating snowline model requires $\kappa_{\mathrm{R}} \approx 4.6~\mathrm{cm^{2}~g^{-1}}$, whereas a self-consistent calculation from the DSHARP opacity yields $\kappa_{\mathrm{R}} \approx 0.4~\mathrm{cm^{2}~g^{-1}}$. 
This mismatch implies excess IR opacity relative to millimeter, possibly arising from a non-MRN size distribution, carbon-rich small grains \citep{WoitkeEtal2016}, or porous dust \citep{ZhangEtal2023}.
\md{Under the assumed dimming curve of luminosity}, our model predicts that the intensity shoulder will move by 10 au over \md{25} years (\fg{sigma_time}), a trend testable with future observations. With ALMA’s growing sample of outburst systems at high resolution, we expect the intensity shoulder to emerge as an ubiquitous feature of large-scale water phase transitions in disks.




\begin{acknowledgements}
    The authors thank for useful discussion with Vardan Elbakyan, Shangjia Zhang, Hanpu Liu. This work is supported by the National Natural Science Foundation of China under grant Nos. 12233004 and 12473065.
\end{acknowledgements}


\bibliographystyle{aa} 
\bibliography{ads} 

\appendix

\section{Bolometric luminosity measurements}
\label{sec:L_bol_uncertain}
The onset of the outburst in V883 Ori was not directly observed. But it likely starts before A.D.~1888, when the reflection nebula IC 430 was observed to be illuminated by the outburst \citep{Pickering1890}, as argued by \citet{StromStrom1993}. 
\md{By fitting the spectral energy distribution, the bolometric luminosity of V883 Ori was estimated to be $400~L_{\odot}$ \citep{StromStrom1993} in the 1980s} \citep{AllenEtal1975,NakajimaEtal1986} and had significantly decreased to $218~L_{\odot}$ \citep{FurlanEtal2016, ConnelleyReipurth2018} in about 20 years. Given the typical month-to-year rise time of FUors (e.g., \citealt{FischerEtal2023}), it is very likely that V883 Ori has been in the dimming phase since A.D.~1888. 

However, estimating the bolometric luminosity of V883 Ori is fraught with substantial uncertainty. As a young stellar object, it suffers severe extinction and reddening at near-infrared wavelengths from dusty proto-stellar envelopes. In the mid-infrared, where extinction is less severe, observations are hindered by heavy saturation in modern surveys like WISE \citep{ContrerasPenaEtal2020}. \citet{LiuEtal2022} derived a much higher bolometric luminosity of ${\approx}647\,L_{\odot}$ by accounting for the reddening. 
\md{Unlike most works adopting discrete photometric data, recently, \citet{CarvalhoEtal2025} fits the medium resolution spectrophotometry within a range of 0.4-2.5 $\mu\mathrm{m}$ and found excellent match to even broader spectrum across 0.4-4.2 $\mu\mathrm{m}$. They yielded an accretion luminosity of 458 $L_{\odot}$, which lies well inside the assumed luminosity curve (\fg{dimming_curve}).}

\section{Dust model}
\label{sec:dust_bins}
At each radius, we model the pebble size distribution by constructing several size bins spanning radii from $s_{\max}$ and $s_{\min}$.
For small pebbles (e.g., $s_{\parti}{<}0.1~\mathrm{cm}$), logarithmic spacing is used, whereas for larger ones, the size bins are more evenly spaced.
The pebble surface density for each size bin ($\Sigma_{\parti,i}$) is then obtained by dividing $\Sigma_{\parti}$ into these size bins according to the MRN distribution $\diff N_{\parti} / \diff s_{\parti} \propto s_{\parti}^{-3.5}$.
For each size bin $i$, its Stokes number ($\mathrm{St}_{i}$) is obtained by scaling the pebble size at the bin center ($s_{\parti,i}$) with $s_{\max}$, assuming that the pebbles follow the Epstein drag law: i.e., $\mathrm{St}_{i,0}= \mathrm{St}_{0} ~s_{\parti,i}/s_{\max}$. Then the 2D pebble density distribution is obtained by adopting normal distribution in the vertical direction:
\begin{equation}
\label{eq:rho_peb}
  \rho_{\parti,i} (z) = \rho_{\parti,0} \exp\left[ -\frac{1}{2} \left( \frac{z}{H_{\parti,i}}\right)^2 \right].
\end{equation}
Here $H_{\parti,i}$ is the scale height of pebbles in the size bin $i$, following \citet{DubrulleEtal1995}:
\begin{equation}
\label{eq:H_peb}
    H_{\parti,i} (z)= H_{\gas} \left[ 1 + \frac{\mathrm{St}_{i,0}}{\alpha}\left(\frac{2 H_{\gas}^{2}}{z^{2}}\right)\left(\exp \left(\frac{z^{2}}{2 H_{\gas}^{2}}\right)-1\right) \right]^{-1/2}.
\end{equation}
In \eq{H_peb}, the vertical variation of Stokes number due to gas stratification is accounted for, $\mathrm{St}_{i}(z) = \mathrm{St}_{i,0} \rho_{\gas}(z)/\rho_{\gas}(0)  \approx \mathrm{St}_{i,0} \exp \left(z^{2}/ 2 H_{\gas}^{2}\right)$. The gas scale height $H_{\gas}$ is determined in the outburst active phase, after iterating between the gas density distribution, assuming hydrostatic balance, and the temperature structure calculated from 2sRT.
Given $\Sigma_{\parti,i}$, the midplane pebble density $\rho_{\parti,0}$ can be obtained by integrating the vertical density distribution following \eq{rho_peb}.

Recently, \citet{HougeEtal2024} preformed dust evolution modeling and demonstrated that, to match the observed spectral index in V883 Ori, pebble sizes should not evolve appreciably during the swift outburst event. Therefore, we maintain a constant pebble size distribution throughout the outburst dimming phase.

Given the size distribution, the total surface area of pebbles per unit volume can be obtained by summing over all size bins,
\begin{equation}
    \label{eq:surface_area}
    \langle \pi s^{2}_{\parti} n_\parti \rangle = \pi\sum_{i} s_{\parti,i}^{2} n_{\parti,i},
\end{equation}
which is used to compute the vapor recondensation rate $\mathcal{R}_{\mathrm{cond}}$. The number density of particles in each size bin is given as $n_{\parti,i} = \rho_{\parti,i}/m_{\parti,i}$, where $m_{\parti,i} = 4\pi \rho_{\bullet}  s_{\parti,i}^{3}/3$ is the particle mass. We adopt $\rho_{\bullet}=1.5~\mathrm{g~cm}^{-3}$ as the internal density of pebbles, assuming a compact ice-silicate mixture with half ice in mass (e.g., \citealt{SchoonenbergOrmel2017}).

\section{Thermal relaxation timescale}
\label{sec:t_relax}
Following \citet{FlockEtal2017}, we combine the optically thin and optically thick limits to determine the thermal relaxation time:
\begin{equation}
\label{eq:t_cool}
    t_{\mathrm{relax}} = t_{\mathrm{thin}} + t_{\mathrm{thick}} = \frac{l_{\mathrm{mfp}}^{2}}{3D_{\mathrm{rad}}} + \frac{H_{\mathrm{phot}}^{2}}{D_{\mathrm{rad}}}
\end{equation}
where $l_{\mathrm{mfp}} = 1/(\kappa_{\mathrm{R}} \rho_{\gas})$ is the mean free path of photons, $D_{\mathrm{rad}} = 16 \sigma_{\mathrm{b}} T^{3} / (3\kappa
_{\mathrm{R}} \rho_{\gas}^{2} C_{\mathrm{v}})$ is the radiation diffusion coefficient and $H_{\mathrm{phot}}$ is the height of disk photosphere ($\tau_{\mathrm{R}}=2/3$), as measured from the simulation. Here $C_{\mathrm{v}} = 8.2\times10^{7}~\mathrm{erg~g^{-1}~K^{-1}}$ is the specific heat capacity of the He-H$_{2}$ mixture (assuming 71\% H$_{2}$ and ideal gas) and $\kappa_{\mathrm{R}}$ is the Rosseland mean opacity per gas mass, which is a free parameter.

\section{2D structure of retreating snowline}
\begin{figure*}
    \centering
    \includegraphics[width=0.9\textwidth]{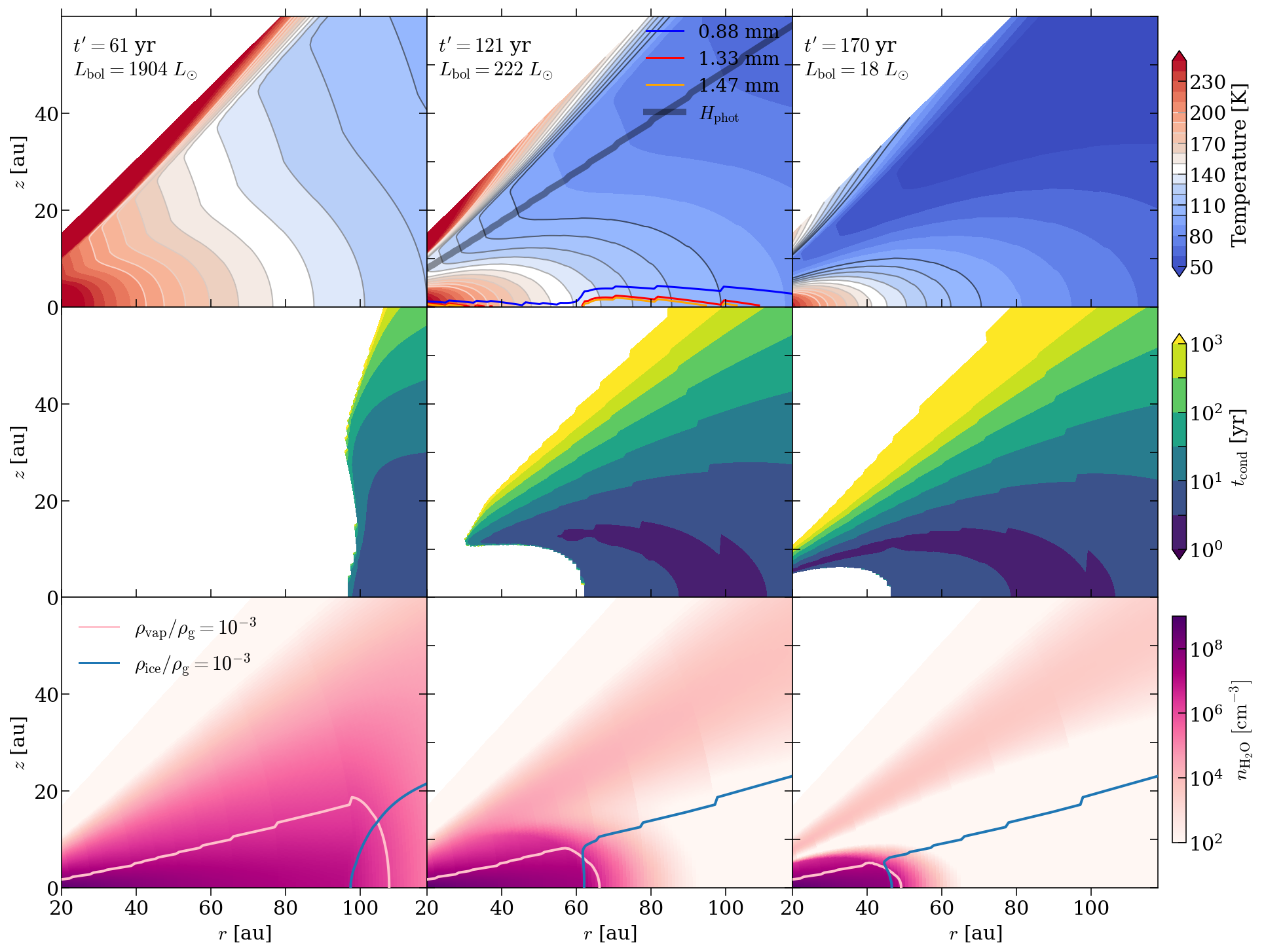}
    \caption{Disk temperature, vapor condensation timescale ($t_{\mathrm{cond}}$) and number density of water ($n_{\mathrm{H_{2}\mathrm{O}}}$) during the outburst dimming phase of the best-fit model. Here $t^{\prime}=t-t_{\mathrm{beg}}$ is the time lapse since the bolometric luminosity started to decline. The value of  $L_{\mathrm{bol}}$ corresponding to this time is labeled.  
    In the top row, the range of 140-150 K is shaded in white to highlight the snowline region. 
    At $t^{\prime}=121~\mathrm{yr}$ -- the present epoch -- the emission surfaces ($\tau_{\mathrm{mm}}=2/3$) of ALMA band 5, 6 and 7 and disk photosphere (see \se{t_relax}) are plotted along with the temperature contour. 
    In the middle row, the region where no condensation occurs is plotted in white. In the bottom row, two lines denoting ice(vapor)-to-gas ratio of $10^{-3}$ are indicated to highlight the snowline region. }
    \label{fig:2D_structure}
\end{figure*}

\label{sec:2D_structure}
To illustrate the evolution of disk's thermal structure during the outburst dimming phase, we present 2D snapshots of our best-fit simulation results at different times since $L_{\mathrm{bol}}$ declines in \fg{2D_structure}. We focus on three epochs: $t^{\prime} = 110~\mathrm{yr}$ (the observational epoch) and two additional snapshots taken 50 yr before (when recondensation just begins) and after (when most vapor has recondensed in the outer disk) this point (see \fg{water_sigma}). At each time, the disk temperature, vapor condensation timescale ($t_{\mathrm{cond}}$) and number density of water vapor ($n_{\mathrm{H}_{2}\mathrm{O}}$) are shown. The condensation timescale is defined as $t_{\mathrm{cond}}=\rho_{\mathrm{vap}}/\mathcal{R}_{\mathrm{cond}}$.

As $L_{\mathrm{bol}}$ decreases, the temperature drops in the entire disk, with the midplane cooling slower due to the large infrared optical depth. This difference in cooling time renders the midplane hotter than the upper layer, which is also found in \citet{LaznevoiEtal2025}.
Concurrently, vapor recondenses and the snowline retreats, as shown in the bottom row of \fg{2D_structure}. 
The middle row of \fg{2D_structure} shows the condensation timescale. In the upper layer of the disk, $t_{\mathrm{cond}}$ is significantly longer than near the midplane, leaving uncondensed vapor plumes in the disk atmosphere (bottom row). This explains the slowly declining tail in the time evolution of the water column density (\fg{water_sigma}). This ``inertia'' in vapor's response to changing luminosity was recently observed in DQ Tau, where the cold water abundance shows little reaction to variations in accretion luminosity on timescales of days to weeks \citep{KospalEtal2025}.
Radially, $t_{\mathrm{cond}}$ reaches a minimum value at ${\sim}80{-}100$ au, increasing outward due to lower pebble densities and inward due to higher temperature. Notably, this increase in condensation timescale was also reported by \citet{SmithEtal2025} in their modeling of EX Lup, implied by the higher water abundance towards the outer disk (their Fig.10).
The ice freezing-out timescale in EX Lup, constrained by multi-epoch observations of \textit{Spitzer} and JWST \citep{SmithEtal2025}, is one magnitude shorter ($\sim 10$ yr) than calculated from our simulations. This difference is expected, given the much larger pebble number density in EX Lup's inner disk.

Finally, we plot the emission surfaces in our 2sRT calculation (upper row, middle panel), where the millimeter optical depth ($\tau_{\mathrm{mm}}$) equals 2/3. There is a clear elevation of the emission surface across snowline at $\approx 60$ au, rising from the recondensation of vapor. As the emission surfaces trace different disk layers thus different temperature, we expect that the continuum emission will not only depend on the surface density distribution but also be modulated by the disk temperature structure. \md{For molecular emission of water at band 5, because $\tau_{\mathrm{mm}} \approx 0.7$ at the midplane of 80 au, we only expect minor dust attenuation on measuring the mean water column density at 80-120 au.}

\section{MCMC analysis}
\label{sec:MCMC_result}

\begin{figure*}
    \centering
    \includegraphics[width=\textwidth]{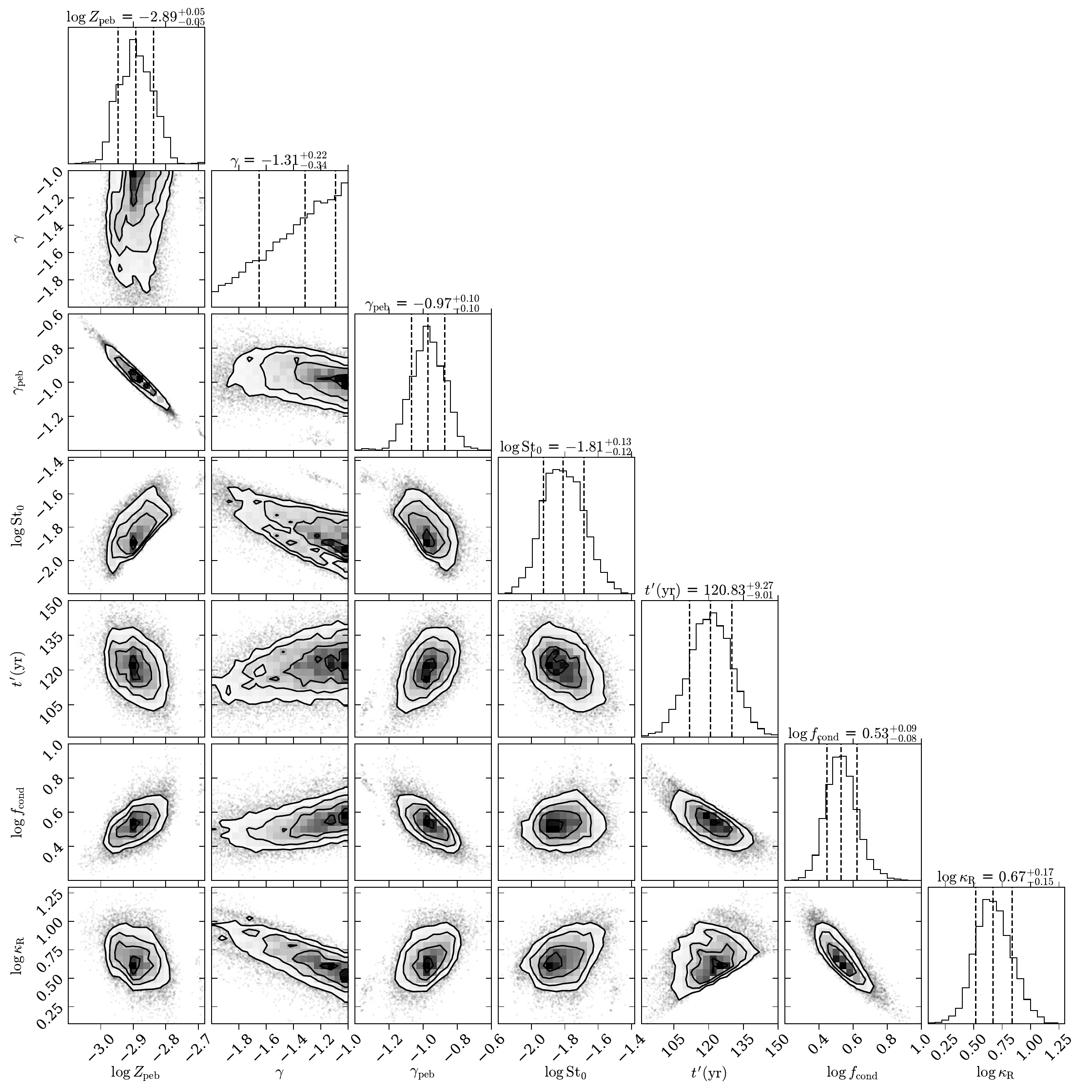}
    \caption{Corner plot of retreating snowline model. The dashed lines indicate 16th, 50th and 84th percentiles of the posterior distribution.}
    \label{fig:corner_outburst}
\end{figure*}

\begin{figure*}
    \centering
    \includegraphics[width=0.8\textwidth]{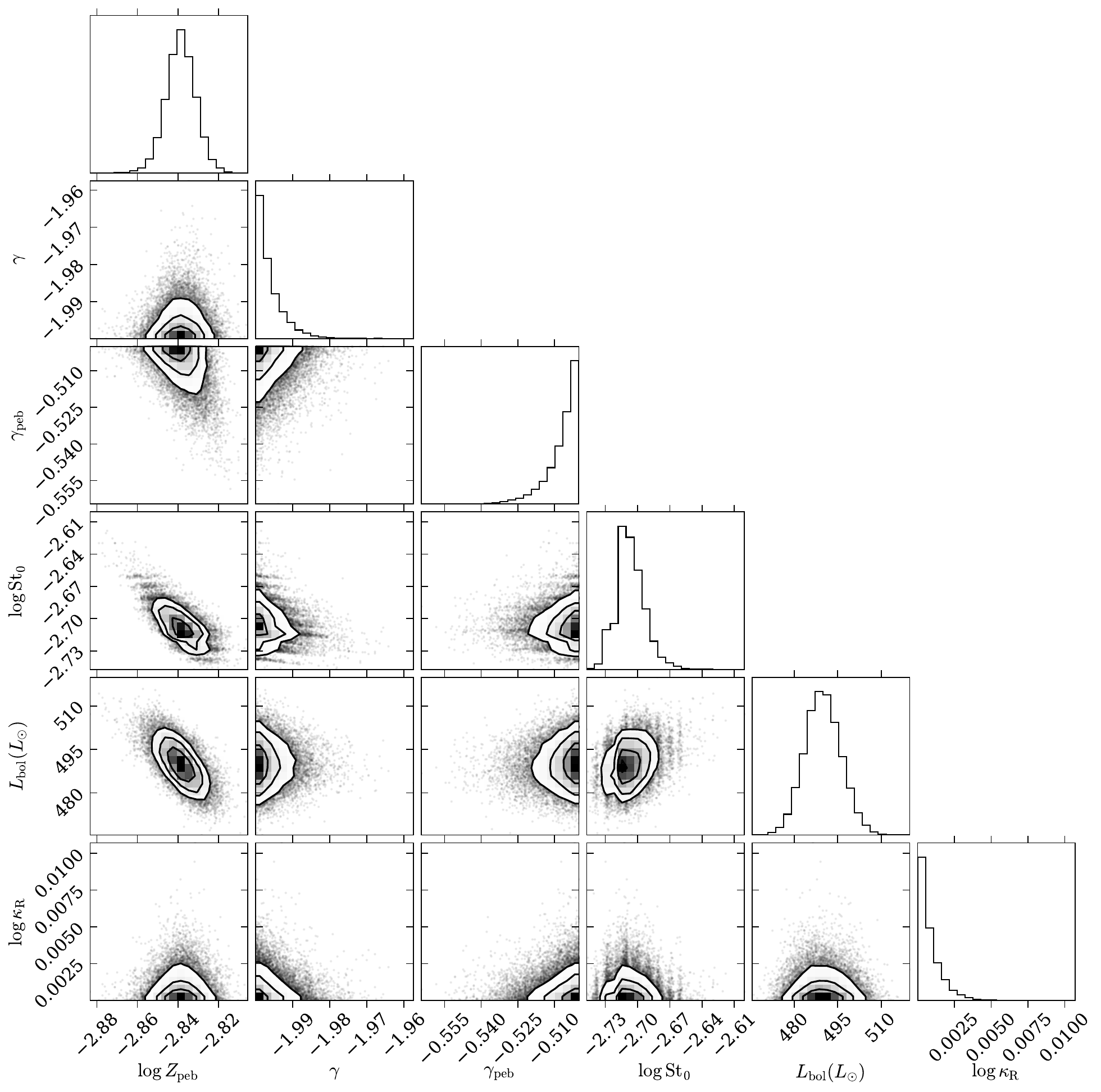}
    \caption{Corner plot of static snowline model.}
    \label{fig:corner_static}
\end{figure*}

To fit the ALMA observations, we adopt a \md{log-normal} likelihood for each dataset:
\begin{equation}
\label{eq:likelihood_i}
    \ln \mathcal{P}_{n}^{i}(x) =  -\frac{1}{2} \left(\frac{\ln x_{n}^{i} - \ln x_{n,\mathrm{syn}}^{i}}{\sigma_{\ln,n}^{i}} \right)^2 - \frac{1}{2} \ln(2 \pi {\sigma^{i}_{\ln,n}}^{2}),
\end{equation}
where $x_{n}^{i}$ is either the azimuthally averaged continuum intensity of the $n$-th \md{annulus} in band $i$ ($I$) or the mean water column density between 80-120 au ($N_{\mathrm{H}_{2}^{18}\mathrm{O}}$). \md{Correspondingly, $x_{n,\mathrm{syn}}^{i}$ is the synthetic data from the simulations.} \md{Following \citet{ViscardiEtal2025},  $\sigma_{\ln,n}^{i} = \sqrt{B_{i}/A_{n}} \times \sigma_{x,n}^{i} / x_{n}^{i}$ refers to the fractional rms error of the mean within annulus $n$, where $B_{i}$ is the beam area of band $i$ and $A_{n}$ is the area corresponding to annulus $n$. For the ${\mathrm{H}_{2}^{18}\mathrm{O}}$ emission, the area spanning 80 to 120 au is treated as a single annulus.}  
Then the joint log-likelihood reads:
\begin{equation}
    \label{eq:likelihood}
    \ln \mathcal{P} = \sum_{i=\mathrm{B6,B7},n} \ln \mathcal{P}_{n}^{i} (I) + \ln \mathcal{P}_{1}^{\mathrm{B5}} (N_{\mathrm{H}_{2}^{18}\mathrm{O}}),
\end{equation}

With this joint likelihood, MCMC analysis is performed for both the retreating snowline model and the static snowline model. In the static model, the parameters $t^{\prime}$ and $f_{\mathrm{cond}}$ are replaced by a single $L_{\mathrm{bol}}$, which controls the temperature structure together with the opacity $\kappa_{\mathrm{R}}$ in the IR band (\se{T_model}). Gaussian priors are used for $t^{\prime}$ ($\sigma_{t^{\prime}} = 10$ yr) and $L_{\mathrm{bol}}$ ($\sigma_{L_{\mathrm{bol}}}=110~L_{\odot}$), assuming that the uncertainties in the luminosity curve span 2$\sigma$ range (i.e., the luminosity curve spans \md{40} yr). Other parameters adopt uniform priors. For $Z_{\mathrm{peb}}$, $\mathrm{St}_{0}$, $f_{\mathrm{cond}}$ and $\kappa_{\mathrm{R}}$, sampling is carried out in logarithmic space.

We run MCMC until all parameters converge over ${>}50$ correlation timescales. 
The resulting posterior distributions are shown in \fgs{corner_outburst}{corner_static}, with values summarized in \tb{mcmc_paras}.
\md{In the retreating snowline model, several correlations can be seen among disk parameters ($Z_{\mathrm{peb}}$, $\gamma$, $\gamma_{\mathrm{peb}}$ and $\mathrm{St}_{0}$) and the gas density slope $\gamma$ remains poorly constrained. These degeneracies are expected since the adopted disk model parameters do not directly correspond to the observational diagnostics. Specifically, the following features are noted:
\begin{enumerate}
    \item The parameter St$_{0}$, which determines the maximum grain size, is positively correlated with $Z_{\mathrm{peb}}$ and negatively correlated with $\gamma$ and $\gamma_{\mathrm{peb}}$. First, the Stokes number of pebbles decreases in denser gas. A shallower (larger) $\gamma$, which implies a higher gas density at 60 au, thus requires a smaller St$_{0}$ to maintain the same $s_{\max}$ of $\sim$ cm, as suggested by the spectral indices (see \se{intensity_shoulder}). Second, pebble size affects the continuum emission differently across the disk. In the inner disk where $s_{\max}\gg \mathrm{mm}$, increasing the pebble size significantly reduces the mm opacity. In the outer disk where $s_{\max} \approx \mathrm{mm}$, however, opacity becomes less sensitive to $s_{\max}$ (e.g., \citealt{BirnstielEtal2018}). Consequently, increasing St$_{0}$ both reduces the overall continuum intensity and flattens its radial slope, necessitating larger $Z_{\mathrm{peb}}$ and a steeper surface density profile (lower $\gamma_{\mathrm{peb}}$) to fit the data.
    \item The gas density slope $\gamma$ is negatively correlated with Rosseland mean opacity $\kappa_{\mathrm{R}}$. This arises since to reproduce the intensity shoulder, the snowline should retreat to $\approx$60 au at the expected observation time ($t^{\prime}$), following the assumed luminosity curve. Because the gas density and $\kappa_{\mathrm{R}}$ together govern the disk's cooling time ($t_{\mathrm{relax}}$), that is, the retreat speed, $\gamma$ and $\kappa_{\mathrm{R}}$ are naturally related. Furthermore, the posterior of $\kappa_{\mathrm{R}}$ essentially follows the assumed Gaussian prior of $t^{\prime}$.
    \item Large $\gamma$ (toward the upper bound) is preferred. This can be understood as follows: during the outburst dimming phase, the outer disk cools faster than the inner regions, steepening the temperature gradient over time \citep{LaznevoiEtal2025}. A higher $\gamma$ (shallower density profile) therefore tends to equalize the cooling rate across the disk, "tilting" the intensity profile towards a plateau, as being observed. A value $\gamma>-1$ may fit the data even better, but such flat profiles are implausible.
    \item The factor $f_{\mathrm{cond}}$, which controls the condensation rate, is negatively correlated with $\kappa_{\mathrm{R}}$. An increase in $\kappa_{\mathrm{R}}$ slows the retreat of the snowline (larger $t^{\prime}$), allowing a reduced $f_{\mathrm{cond}}$ to achieve similar condensation levels.
\end{enumerate}
From \fg{obv_compare}, we find that the model underestimates the emission within 40 au and correspondingly, struggles to capture the declining spectral index towards inner disk. These discrepancies likely arise from intense viscous heating \citep{AlarconEtal2024} and optically thick environment \citep{TobinEtal2023} in V883 Ori's inner disk., which is included in the model.
Simple power-law profiles, as adopted in this work, cannot simultaneously reproduce both the inner and outer disk, implying a sharp transition in the disk properties of V883 Ori.
}

In the static snowline model, many parameters are driven toward the edges of their prior ranges, indicating that the model cannot adequately reproduce the data. In particular, the disk struggles to reach sufficiently high temperatures. Consequently, $\kappa_{\mathrm{R}}$ \md{and $\gamma$} are pushed to \md{their} minimum value, and $L_{\mathrm{bol}}$ is constrained to the upper wing of its Gaussian prior, trying to obtain maximum penetration of stellar photons into the disk midplane regions.

\end{document}